# Desensitized Cubature Kalman Filter with Uncertain Parameter

**Taishan Lou**

*School of Electric and Information Engineering, Zhengzhou University of Light Industry,
Zhengzhou, 45002, China (e-mail: tayzan@sina.com)*

**Abstract:** A robust desensitized cubature Kalman filtering (DCKF) for nonlinear systems with uncertain parameter is proposed. Sensitivity matrices are defined as the integral form, and desensitized cost function is designed by penalizing the posterior covariance trace by a sensitivity-weighting sum of the posteriori sensitivities. The DCKF gain is obtained by minimizing the desensitized cost function to amend the state estimation. Then, the sensitivity propagation of the state estimate errors is described, and the sensitivity of the root square matrix is obtained by solving a special equation. The effectiveness of the proposed DCKF was demonstrated by two numerical simulations with uncertain parameters.

## 1. INTRODUCTION

Nonlinear state estimation plays an important role in a wide variety of applications, such as target tracking, navigation for the aerospace vehicle, chemical plant control, multi-sensor data fusion, etc. Many nonlinear filters based on the Bayesian framework have been put forward in over four decades. One inevasible difficulty of the application of the Bayesian estimation to the practice problems is that the realistic dynamic systems are often nonlinear [1, 2]. These nonlinear filters can be roughly classified as three categories. The first one is the filters based on the analytical approximation approach (also called function approximation approach), such as the extended Kalman filter (EKF) and the second-order EKF [3]. The second one is the filters based on the deterministic sampling approach (also called sigma point Kalman filter), which contain the unscented Kalman filter (UKF) [4], divided difference filter [4], and Gauss-Hermite quadrature filter [6]. The third one is the filters based on the Monte Carlo simulation, such as sequential Monte Carlo [7] and particle filter [8]. For the first category, calculating Jacobians or Hessians are often numerically unstable and computationally intensive in the filters. The nonlinear filters in the second category suffer from the curse of dimensionality or divergence or both [2]. For the third category, the filters may solve any non-Gaussian estimation problems, but the computational load is often expensive [1].

Recently, a nonlinear filter based on the Bayesian framework, which has been proposed for the nonlinear Gaussian systems, is the cubature Kalman filter (CKF) [9, 10]. The core idea of the CKF is using the cubature role to approximate the multidimensional integrals with 2n deterministic cubature points, where n is the state vector dimension. In the CKF algorithm, the mean and covariance of the state and measurement vectors are calculated by propagating the cubature points through the nonlinear function. Comparison of the EKF and the UKF, the CKF is usually applied in the high-dimensional systems, and has more accuracy and numerical stability [9]. Subsequently, the CKF is used in many applications, such as navigation [11], sensor data fusion [11], etc.

In these filters above, there is a fundamental assumption that the dynamic models can be accurately modeled without any unknown statistical properties of the noises, or uncertain parameters [3]. When the assumption is satisfied, the performance of these filters will be highly sensitive to the dynamic model uncertainties, and deteriorate appreciably [3]. A desensitized optimal control methodology is extended by Karlgaard and Shen [14] to the robust filter design problem that the performance sensitivity of the filters respected to model uncertain parameters can be reduced. The desensitized Kalman filter is designed by penalizing the cost function consisting of the posterior covariance trace using a weighted norm of the state error sensitivities, and minimizing this cost function to obtain the desensitized state estimates [14]. Subsequently, the desensitized divided difference filtering [14] and the desensitized unscented Kalman filtering (DUKF) [16] are presented. For the DUKF, there is no continuous propagation for the sensitivities of the sigma points between the iterations, because a new set of sigma points is always resampled at the next iteration. Shen and Karlgaard [16] skillfully designed a unique way to propagate the sensitivities of the state estimate error and the a priori/posteriori covariance matrices for the DUKF.

This paper proposes a robust cubature Kalman filter for systems with uncertain parameter. A desensitized cost function of the robust desensitized cubature Kalman filtering (DCKF) is designed by penalizing the a posteriori covariance trace by a sensitivity-weighting sum of the a posteriori sensitivities. Then, a gain matrix of the DCKF is obtained by minimizing the new cost function to amend the state estimation. Section 2 briefly introduces the recursive Bayesian framework and the CKF algorithm. Section 3 presents the propagation of the sensitivities and the DCKF is proposed naturally. Two numerical simulations about a vertically falling body model and a hovering helicopter model (HHM) are analyzed in Section 4. The conclusions are summarized in Section 5.

## 2. CUBATURE KALMAN FILTER

Consider the discrete nonlinear process and measurement models with additive noises given by

$$x_k = f(x_{k-1}, c, u_{k-1}) + w_{k-1} \quad (1)$$

$$z_k = h(x_k, c, u_k) + v_k \quad (2)$$

where $x_k$ is the $n \times 1$ state vector, and $z_k$ is the $m \times 1$ measurement vector. $f$ is the dynamic vector-valued function, $h$ is the nonlinear measurement vector-valued function. $c$ is referred to as the uncertain parameter vector. $u_k$ is the known control input vector. $w_k$ and $v_k$ are independent zero-mean Gaussian noise processes, and their covariance are respectively $Q_k$ and $R_k$. They satisfy

$$E[w_k w_j^T] = Q_k \delta_{ij}, \; E[v_k v_j^T] = R_k \delta_{ij}, \; E[w_k v_j^T] = 0 \quad (3)$$

where $\delta_{kj}$ is the Kronecker delta function, and $Q_k$ is positive semi-definite, and $R_k$ is positive definite.

### 2.1 Recursive Bayesian filter under Gaussian domain

Under the Gaussian domain, the predictive density $p(x_k|Z_{k-1})$, the filter likelihood density $p(z_k|Z_k)$ and the a posterior density $p(x_k|Z_k)$ are both Gaussian. When the a posterior density $p(x_{k-1}|Z_{k-1}) = N(\hat{x}_{k-1}^+, P_{k-1}^+)$ at time $t_{k-1}$ and their distributions respectively are

$$\begin{aligned} p(x_k|Z_{k-1}) &= N(\hat{x}_k^-, P_k^-) \\ p(z_k|Z_{k-1}) &= N(\hat{z}_k^-, P_{zz,k}^-) \\ p(x_k|Z_k) &= N(\hat{x}_k^+, P_k^+) \end{aligned} \quad (4)$$

where $Z_k = \{u_i, z_i\}_{i=1}^k$ is the history of input measurement pairs up to time $t_k$, $N(\cdot, \cdot)$ is the Gaussian density symbol, the superscripts "$-$" denote a priori and "$+$" denote a posteriori. $\hat{x}_k^-$, $\hat{x}_k^+$, $\hat{z}_k^-$ are the state estimates and the measurement estimate, respectively; $P_k^-$, $P_k^+$, $P_{zz,k}^-$ are the corresponding error covariance matrices, respectively.

Then, the functional recursion of the Bayesian filter reduces to an algebraic recursion. Only the means and covariances of various conditional densities encountered in the time and measurement updates are needed to calculate. The recursive Bayesian filter under the Gaussian approximation is summarized as follows [9]:

Time update

$$\begin{aligned} \hat{x}_k^- &= E[x_k|Z_{k-1}] = E[f(x_{k-1}, \bar{c}, u_{k-1}) + w_{k-1}|Z_{k-1}] \\ &= E[f(x_{k-1}, \bar{c}, u_{k-1})|Z_{k-1}] \\ &= \int_{R^n} f(x_{k-1}, \bar{c}, u_{k-1}) p(x_{k-1}|Z_{k-1}) dx_{k-1} \\ &= \int_{R^n} f(x_{k-1}, \bar{c}, u_{k-1}) N(\hat{x}_{k-1}^+, P_{k-1}^+) dx_{k-1} \end{aligned} \quad (5)$$

$$\begin{aligned} P_k^- &= E[\tilde{x}_k^- \tilde{x}_k^{-T} | z_{1:k-1}] \\ &= \int_{R^n} f(x_{k-1}, \bar{c}, u_{k-1}) f^T(x_{k-1}, \bar{c}, u_{k-1}) N(\hat{x}_{k-1}^+, P_{k-1}^+) dx_{k-1} \\ &\quad - \hat{x}_k^- \hat{x}_k^{-T} + Q_{k-1} \end{aligned} \quad (6)$$

where $\tilde{x}_k^- = \hat{x}_k^- - x_k$ is the a priori estimation error, and $\bar{c}$ is the reference value of the parameter vector $c$.

Measurement update

$$\hat{z}_k^- = E[z_k|Z_{k-1}] = \int_{R^n} h(x_k, \bar{c}, u_k) N(\hat{x}_k^-, P_k^-) dx_k \quad (7)$$

$$\begin{aligned} P_{zz,k}^- &= \int_{R^n} h(x_k, \bar{c}, u_k) h^T(x_k, \bar{c}, u_k) N(\hat{x}_k^-, P_k^-) dx_{k-1} \\ &\quad - \hat{z}_k^- \hat{z}_k^{-T} + R_k \end{aligned} \quad (8)$$

$$P_{xz,k}^- = \int_{R^n} x_k h^T(x_k, c, u_k) N(\hat{x}_k^-, P_k^-) dx_{k-1} - \hat{x}_k^- \hat{z}_k^{-T} \quad (9)$$

$$\hat{x}_k^+ = \hat{x}_k^- + K_k [z_k - \hat{z}_k^-] \quad (10)$$

$$P_k^+ = E[\tilde{x}_k^+ \tilde{x}_k^{+T}] = P_k^- - P_{xz,k}^- K_k^T - K_k P_{xz,k}^{-T} + K_k P_{zz,k}^- K_k^T \quad (11)$$

$$K_k = P_{xz,k}^- (P_{zz,k}^-)^{-1} \quad (12)$$

where $\tilde{x}_k^+ = \hat{x}_k^+ - x_k$ is the a posteriori estimation error.

The Kalman optimal gain $K_k$ is obtained by minimizing the cost function $J = Tr(P_k^+)$, in which "$Tr$" denotes the trace of the matrix, and the result is (12).

### 2.2 Cubature Kalman filter algorithm

The cubature Kalman filter is a type of Bayesian filter under Gaussian assumption. The CKF approximates the mean and covariance of a random variable propagated under a nonlinear function by the cubature rule. Under the Gaussian assumption, the functional recursion of the Bayesian filter reduces to an algebraic recursion, in which the multidimensional integrals can be approximated by the cubature rule [9, 17].

The cubature rule approximates the $n$-dimensional Gaussian weighted integral by using the mean $\mu$ and covariance $P$. The formulation is

$$\int_{R^n} f(x) N(\mu, P) dx \approx \frac{1}{2n} \sum_{i=1}^{2n} f(\mu + \sqrt{P} \xi_i) \quad (13)$$

where $\sqrt{P}$ is a square-root factor of the covariance $P$, which satisfy $P = \sqrt{P}(\sqrt{P})^T$; $\xi_i$ is the $i^{th}$ element of $2n$ cubature points, and its value comes from the following set

$$\sqrt{n} \{e_1, e_2, \cdots, e_n, -e_1, -e_2, \cdots, -e_n\} \quad (14)$$

where $e_i (i=1, 2, \cdots, n)$ is the unit vector, in which the $i^{th}$ element is one and others are zeros, $e_i = [0, \cdots, 0, 1, 0, \cdots, 0]^T$.

Based on the state estimate $\hat{x}_{k-1}^+$ and covariance at time $t_{k-1}$, the cubature points are generated as follows

$$\chi_{j,k-1}^+ = \sqrt{P_{k-1}^+} \xi_j + \hat{x}_{k-1}^+, j = 1, 2, \cdots, 2n \quad (15)$$

Each of the propagated cubature points is computed through the nonlinear function as

$$\chi_{j,k}^{*-} = f(\chi_{j,k-1}^+, \bar{c}, u_{k-1}), j = 1, 2, \cdots, 2n \quad (16)$$

Then, the a priori state and the corresponding error covariance in (5) and (6) are evaluated as

$$\hat{x}_k^- = \frac{1}{2n}\sum_{j=1}^{2n}\chi_{j,k}^{*-} \quad (17)$$

$$P_k^- = \frac{1}{2n}\sum_{j=1}^{2n}\chi_{j,k}^{*-}\chi_{j,k}^{*-T} - \hat{x}_k^-\hat{x}_k^{-T} + Q_{k-1} \quad (18)$$

Redraw the cubature points by using $\hat{x}_k^-$ and $P_k^-$

$$\chi_{j,k}^- = \sqrt{P_k^-}\xi_j + \hat{x}_k^-, j=1,2,\cdots,2n \quad (19)$$

Then, by using (7) to (9), the predicted measurement and its corresponding covariances are

$$Z_{j,k}^- = h(\chi_{j,k}^-, \bar{c}, u_k) \quad (20)$$

$$\hat{z}_k^- = \frac{1}{2n}\sum_{j=1}^{2n}Z_{j,k}^- \quad (21)$$

$$P_{zz,k}^- = \frac{1}{2n}\sum_{j=1}^{2n}Z_{j,k}^- Z_{j,k}^{-T} - \hat{z}_k^-\hat{z}_k^{-T} + R_k \quad (22)$$

$$P_{xz,k}^- = \frac{1}{2n}\sum_{j=1}^{2n}\chi_{j,k}^- Z_{j,k}^{-T} - \hat{x}_k^-\hat{z}_k^{-T} \quad (23)$$

Finally, the a posterior estimate and the associated covariance are given by

$$\hat{x}_k^+ = \hat{x}_k^- + K_k(z_k - \hat{z}_k^-) \quad (24)$$

$$P_k^+ = P_k^- - P_{xz,k}^- K_k^T - K_k P_{xz,k}^{-T} + K_k P_{zz,k}^- K_k^T \quad (25)$$

$$K_k = P_{xz,k}^-(P_{zz,k}^-)^{-1} \quad (26)$$

where the CKF gain $K_k$ is obtained by minimizing its cost function $J_c = Tr(P_k^+)$.

## 3. DESENSITIZED CUBATURE KALMAN FILTER

In this section, the robust DCKF is presented by using the Sensitivity propagation equation and the CKF method in section 2. The difficulty for the DCKF compared with the DEKF is the propagation of the sensitivities, because the cubature points are always redrawn time after time in the filtering, and this makes that there is no continuity about the cubature points from one iteration to the next. So, the propagation of the sensitivities is introduced firstly as in the literature [16], and the DCKF algorithm is summarized in table 2.

### 3.1 Sensitivity propagation equation for the recursive Bayesian filter

Under the basic assumptions of the recursive Bayesian filter, such as no model and parameter uncertainties, the state estimate are unbiased. It means that the optimal estimation errors satisfy

$$E[\tilde{x}_k^-] = 0, E[\tilde{x}_k^+] = 0 \quad (27)$$

When the model parameter has uncertainty, the basic assumptions of the recursive Bayesian filter cannot be satisfied, and the state estimates may be biased and even divergence. So, the state estimate error sensitivities of each parameter, $c_i$, could be defined as [14]

$$s_{i,k}^- = \frac{\partial \tilde{x}_k^-}{\partial c_i} = \frac{\partial \hat{x}_k^-}{\partial c_i} \quad (28)$$

$$s_{i,k}^+ = \frac{\partial \tilde{x}_k^+}{\partial c_i} = \frac{\partial \hat{x}_k^+}{\partial c_i} \quad (29)$$

where $c_i$ denotes the $i^{th}$ component of the parameter vector. Note that the sensitivity of the true state is $\partial x/\partial c_i = 0$ in (28) and (29) because the true state doesn't vary with the assumed value of the parameter vector $c$.

The propagation equations of the recursive Bayesian filter are defined as

$$s_{i,k}^- \square E[\frac{\partial f(x_{k-1}, c, u_{k-1})}{\partial c_i}]$$
$$= \int_{R^n} \frac{\partial f(x_{k-1}, c, u_{k-1})}{\partial c_i} N(\hat{x}_{k-1}^+, P_{k-1}^+) dx_{k-1} \quad (30)$$

$$s_{i,k}^+ = s_{i,k}^- - K_k\gamma_{i,k} \quad (31)$$

where the sensitivity of the nonlinear measurement function is defined by

$$\gamma_{i,k} \square E[\frac{\partial h(x_k, c, u_k)}{\partial c_i}] = \int_{R^n} \frac{\partial h(x_k, c, u_k)}{\partial c_i} N(\hat{x}_k^-, P_k^-) dx_k \quad (32)$$

Note that the sensitivity of gain matrix is assumed as $\partial K/\partial c_i = 0$ in (31), because any $\partial K/\partial c_i \neq 0$ means that the solution for the optimal gain is a function of the residual, which violates the basis for the linear update equation given in (11) [14].

### 3.2 Sensitivity propagation of cubature points

The propagation of the sensitivities bases on the sensitivity of cubature points, which is obtained by taking partial derivative of the cubature point generated equation, such as (15) and (19). The propagation equations of the sensitivities can be obtained by taking partial derivative of the propagation equations of the CKF, such as (16)-(18) and (20)-(25). The sensitivity propagation algorithm of the cubature points is summarized in Table 1.

**Table 1. Sensitivity propagation algorithm of cubature points**

| Time update |
| --- |
| (1) Compute the sensitivities of step $k-1$ cubature points ($i=1,2,\cdots,\ell, j=1,2,\cdots,2n$) $$\frac{\partial \chi_{j,k-1}^+}{\partial c_i} = \frac{\partial \sqrt{P_{k-1}^+}}{\partial c_i}\xi_j + s_{i,k-1}^+ \quad (33)$$ |
| (2) Propagate the sensitivity cubature points $$\frac{\partial \chi_{j,k}^{*-}}{\partial c_i} = \frac{\partial f(\chi_{j,k-1}^+, \bar{c}, u_{k-1})}{\partial \chi_{j,k-1}^+}\frac{\partial \chi_{j,k-1}^+}{\partial c_i} + \frac{\partial f(\chi_{j,k-1}^+, \bar{c}, u_{k-1})}{\partial c_i} \quad (34)$$ |
| (3) Evaluate the sensitivities of the prior state estimate $$s_{i,k}^- = \frac{\partial \hat{x}_k^-}{\partial c_i} = \frac{1}{2n}\sum_{j=1}^{2n}\frac{\partial \chi_{j,k}^{*-}}{\partial c_i} \quad (35)$$ |
| (4) Evaluate the sensitivities of the prior covariance matrix |
| $$\frac{\partial \boldsymbol{P}_k^-}{\partial c_i} = \frac{1}{2n}\sum_{j=1}^{2n}\left\{\frac{\partial \boldsymbol{\chi}_{j,k}^{*-}}{\partial c_i}\boldsymbol{\chi}_{j,k}^{*-T} + \boldsymbol{\chi}_{j,k}^{*-}\left(\frac{\partial \boldsymbol{\chi}_{j,k}^{*-}}{\partial c_i}\right)^T\right\} \quad (36)$$ $$- \boldsymbol{s}_{i,k}^-\hat{\boldsymbol{x}}_k^{-T} - \hat{\boldsymbol{x}}_k^-\boldsymbol{s}_{i,k}^{-T}$$ |
| **Measurement update** |
| (5) Compute the sensitivities of the redrawn cubature points $$\frac{\partial \boldsymbol{\chi}_{j,k}^-}{\partial c_i} = \frac{\partial \sqrt{\boldsymbol{P}_k^-}}{\partial c_i}\boldsymbol{\xi}_i + \boldsymbol{s}_{i,k}^- \quad (37)$$ |
| (6) Evaluate the sensitivity of the predicted measurement cubature points $$\frac{\partial \boldsymbol{Z}_{j,k}^-}{\partial c_i} = \frac{\partial h(\boldsymbol{\chi}_{j,k}^-,\overline{\boldsymbol{c}},\boldsymbol{u}_k)}{\partial \boldsymbol{\chi}_{j,k}^-}\frac{\partial \boldsymbol{\chi}_{j,k}^-}{\partial c_i} + \frac{\partial h(\boldsymbol{\chi}_{j,k}^-,\overline{\boldsymbol{c}},\boldsymbol{u}_k)}{\partial c_i} \quad (38)$$ |
| (7) Evaluate the sensitivities of the predicted measurement $$\boldsymbol{\gamma}_{i,k} = \frac{\partial \hat{\boldsymbol{z}}_k^-}{\partial c_i} = \frac{1}{2n}\sum_{j=1}^{2n}\frac{\partial \boldsymbol{Z}_{j,k}^-}{\partial c_i} \quad (39)$$ |
| (8) Evaluate the sensitivities of the innovation covariance matrix $$\frac{\partial \boldsymbol{P}_{zz,k}^-}{\partial c_i} = \frac{1}{2n}\sum_{j=1}^{2n}\left\{\frac{\partial \boldsymbol{Z}_{j,k}^-}{\partial c_i}\boldsymbol{Z}_{j,k}^{-T} + \boldsymbol{Z}_{j,k}^-\left(\frac{\partial \boldsymbol{Z}_{j,k}^-}{\partial c_i}\right)^T\right\} \quad (40)$$ $$- \boldsymbol{\gamma}_{i,k}\hat{\boldsymbol{z}}_k^{-T} - \hat{\boldsymbol{z}}_k^-\boldsymbol{\gamma}_{i,k}^T$$ |
| (9) Evaluate the sensitivities of the cross-covariance matrix $$\frac{\partial \boldsymbol{P}_{xz,k}^-}{\partial c_i} = \frac{1}{2n}\sum_{j=1}^{2n}\left\{\frac{\partial \boldsymbol{\chi}_{j,k}^-}{\partial c_i}\boldsymbol{Z}_{j,k}^{-T} + \boldsymbol{\chi}_{j,k}^-\left(\frac{\partial \boldsymbol{Z}_{j,k}^-}{\partial c_i}\right)^T\right\} \quad (41)$$ $$- \boldsymbol{s}_{i,k}^-\hat{\boldsymbol{z}}_k^{-T} - \hat{\boldsymbol{x}}_k^-\boldsymbol{\gamma}_{i,k}^T$$ |
| (10) Evaluate the sensitivities of the posterior state estimate $$\boldsymbol{s}_{i,k}^+ = \boldsymbol{s}_{i,k}^- - \boldsymbol{K}_k\boldsymbol{\gamma}_{i,k} \quad (42)$$ |
| (11) Evaluate the sensitivities of the posterior covariance matrix $$\frac{\partial \boldsymbol{P}_k^+}{\partial c_i} = \frac{\partial \boldsymbol{P}_k^-}{\partial c_i} - \frac{\partial \boldsymbol{P}_{xz,k}^-}{\partial c_i}\boldsymbol{K}_k^T - \boldsymbol{K}_k\frac{\partial \boldsymbol{P}_{xz,k}^{-T}}{\partial c_i} \quad (43)$$ $$+ \boldsymbol{K}_k\frac{\partial \boldsymbol{P}_{zz,k}^-}{\partial c_i}\boldsymbol{K}_k^T$$ |

The sensitivity of the root square matrix $\sqrt{\boldsymbol{P}}$ must be computed when the sensitivities of the cubature points and redrawn cubature points are calculated. By taking partial derivative of equation $\boldsymbol{P}=\sqrt{\boldsymbol{P}}(\sqrt{\boldsymbol{P}})^T$, which is different from the equation in Shen and Karlgaard [16], they satisfy equation

$$\frac{\partial \boldsymbol{P}}{\partial c_i} = \frac{\partial \sqrt{\boldsymbol{P}}}{\partial c_i}(\sqrt{\boldsymbol{P}})^T + \sqrt{\boldsymbol{P}}(\frac{\partial \sqrt{\boldsymbol{P}}}{\partial c_i})^T \quad (44)$$

The solution $\partial\sqrt{\boldsymbol{P}}/\partial c_i$ of (44) can be obtained as [18]

$$\frac{\partial \sqrt{\boldsymbol{P}}}{\partial c_i} = \Psi^{-1}\left\{\frac{1}{2}\Psi\left(\frac{\partial \boldsymbol{P}}{\partial c_i}\right)^T\Psi^T - \Gamma\right\}^{-1}\Theta^T \quad (45)$$

where $\Gamma$ is an arbitrary $n\times n$ skew symmetric matrix which satisfies $\Gamma^T=-\Gamma$, $\Psi$ and $\Theta$ are non-singular matrix such that $\Psi\sqrt{\boldsymbol{P}}\Theta=\boldsymbol{I}$.

*3.3 Desensitized cubature Kalman filter algorithm*

Under the recursive Bayesian filter framework, the CKF is introduced in the above section. The DCKF is naturally obtained based on the CKF by using the sensitivity propagation of cubature points.

With the sensitivities of the posterior state estimate in Table 1, a desensitized cost function of the DCKF, which consists of the posterior covariance and a weighted norm of the posterior sensitivity, is penalized by a sensitivity-weighting sum of the sensitivities $\boldsymbol{s}_{i,k}^+$

$$J_d = Tr(\boldsymbol{P}_k^+) + \sum_{i=1}^{\ell}\boldsymbol{s}_{i,k}^{+T}\boldsymbol{W}_{i,k}\boldsymbol{s}_{i,k}^+ \quad (46)$$

where $\boldsymbol{W}_{i,k}$ is a $n\times n$ symmetric positive semi-definite weighting matrix for the $i^{th}$ sensitivity.

Substituting (25) and the sensitivities of the posterior state estimate in Table 1 into (46), taking partial derivative of $J_d$ with respected to the gain matrix $\boldsymbol{K}_k$, and setting the partial derivative $\partial J_d/\partial \boldsymbol{K}_k=0$ gives the solution

$$\boldsymbol{K}_k\boldsymbol{P}_{zz,k}^- + \sum_{i=1}^{\ell}\boldsymbol{W}_{i,k}\boldsymbol{K}_k\boldsymbol{\gamma}_{i,k}\boldsymbol{\gamma}_{i,k}^T = \boldsymbol{P}_{xz,k}^- + \sum_{i=1}^{\ell}\boldsymbol{W}_{i,k}\boldsymbol{s}_{i,k}^-\boldsymbol{\gamma}_{i,k}^T \quad (47)$$

Also, the gain $\boldsymbol{K}_k$ is obtained by algebraically solving with the linear equation in (47).

The DCKF algorithm is summarized in Table 2.

**Table 2. DCKF algorithm**

| |
|---|
| **Time update** |
| (1) Initialize the state vector, $\hat{\boldsymbol{x}}_0$, the auxiliary matrix, $\boldsymbol{P}_0$, and the sensitivity parameters, $\boldsymbol{s}_0$ and $\partial \boldsymbol{P}_0/\partial c_i, i=1,2,\cdots,\ell$. |
| (2) Factorize $$\boldsymbol{P}_{k-1}^+ = \sqrt{\boldsymbol{P}_{k-1}^+}(\sqrt{\boldsymbol{P}_{k-1}^+})^T \quad (48)$$ |
| (3) Evaluate the cubature points, $\boldsymbol{\chi}_{j,k-1}^+$, using (15), and the sensitivities, $\partial\boldsymbol{\chi}_{j,k-1}^+/\partial c_i$, using (45) and (33). |
| (4) Evaluate the propagated cubature points, $\boldsymbol{\chi}_{j,k}^{*-}$, using (16), and the sensitivities, $\partial\boldsymbol{\chi}_{j,k}^{*-}/\partial c_i$, using (34). |
| (5) Estimate the predicted state, $\hat{\boldsymbol{x}}_k^-$, using (17), and the prior sensitivity, $\boldsymbol{s}_{i,k}^-$, using (35). |
| (6) Estimate the predicted covariance matrix, $\boldsymbol{P}_k^-$, using (18), and the sensitivity, $\partial\boldsymbol{P}_k^-/\partial c_i$, using (36). |
| **Measurement update** |
| (7) Factorize $$\boldsymbol{P}_k^- = \sqrt{\boldsymbol{P}_k^-}(\sqrt{\boldsymbol{P}_k^-})^T \quad (49)$$ |
| (8) Evaluate the redrawn cubature points, $\boldsymbol{\chi}_{j,k}^-$, using (19), and the sensitivities, $\partial\boldsymbol{\chi}_{j,k-1}^+/\partial c_i$, using (45) |
| and (37). |
| (9) Evaluate the propagated cubature points of measurement equation, $Z_{j,k}^-$, using (20), and the sensitivities, $\partial Z_{j,k}^-/\partial c_i$, using (38). |
| (10) Estimate the predicted measurement, $\hat{z}_k^-$, using (21), and the sensitivity, $\gamma_{i,k}$, using (39). |
| (11) Estimate the innovation covariance matrix, $P_{zz,k}^-$, using (22), and the sensitivity, $\partial P_{zz,k}^-/\partial c_i$, using (40). |
| (12) Estimate the cross-covariance matrix, $P_{xz,k}^-$, using (23), and the sensitivity, $\partial P_{xz,k}^-/\partial c_i$, using (41). |
| (13) Obtain the gain matrix, $K_k$, by solving (47). |
| (14) Estimate the posterior state, $\hat{x}_k^+$, using (24), and the posterior sensitivity, $s_{i,k}^+$, using (42). |
| (15) Estimate the posterior covariance matrix, $P_k^+$, using (25), and the sensitivity, $\partial P_k^+/\partial c_i$, using (43). |

## 4. NUMERICAL SIMULATION

To verify the effectiveness of the proposed DCKF algorithm in the previous section, the vertical falling body model with one uncertain parameter [18] and the HHM with two uncertain parameters are considered [21]. The perfect CKF (perf. CKF) and the imperfect CKF (imp. CKF) are employed to compare the performance of the proposed algorithm. The "perfect" means that the true values of the parameters are all known exactly in perfect CKF; The "imperfect" means that in the imperfect CKF the true values of the uncertain parameters are not known, and only the reference values, coming from the previous experience, are known.

### 4.1 Falling body model with one uncertain parameter

In this example, we use the perf. CKF, the imp. CKF and the proposed DCKF to estimate the altitude $x_1(t)$, velocity $x_2(t)$ and constant ballistic coefficient $x_3(t)$ of a vertically falling body with a high velocity [19, 20, 22]. A tracking radar device is used to record the range measurements between the radar and the falling body. The system equations are given by

$$\dot{x}_1(t) = x_2(t) + w_1(t)$$
$$\dot{x}_1(t) = x_2^2(t)x_3(t)\exp\{-x_1(t)/c\} - g + w_2(t) \quad (50)$$
$$\dot{x}_3(t) = w_3(t)$$

where $c$ is a constant which is the relationship between the air density and the altitude and it is reference value is $\bar{c} = 2\times 10^4$, $g = 32.2\,\text{ft/s}^2$ is the gravitational acceleration. $w_i(t)$ is the process noise that affects the $i^{\text{th}}$ equation with $E[w_i^2(t)] = 0\ (i=1,2,3)$. The discrete-time range measurement from the radar is given by

$$z_k = \sqrt{M^2 + (x_{1,k} - H)^2} + v_k \quad (51)$$

where $M = 10^5\,\text{ft}$ is the radar horizontal distance from the body's vertical line of fall, and $H = 10^5\,\text{ft}$ is the radar altitude above the ground level. $v_k$ is the measurement noise assumed to be a zero-mean Gaussian noise with covariance $E[v_k^2] = R = 10^4\,\text{ft}^2$.

Here, we assumed that there is uncertainties in the constant parameter $c$, and the true values of $c$ subjects to a uniform distribution, $c \sim U(3/4\bar{c}, 5/4\bar{c})$. The true state and initial estimates are given as

$$x_1(0) = 3\times 10^5\,\text{ft};\ x_2(0) = -2\times 10^4\,\text{ft/s};\ x_3(0) = 1\times 10^{-3} \quad (52)$$
$$\hat{x}_1(0) = 3\times 10^5\,\text{ft};\ \hat{x}_2(0) = -2\times 10^4\,\text{ft/s};\ \hat{x}_3(0) = 3\times 10^{-5} \quad (53)$$

and the initial covariance is given as

$$P_0^+ = diag\{1\times 10^6\,\text{ft}^2, 4\times 10^6\,\text{ft}^2/s^2, 1\times 10^{-4}\} \quad (54)$$

Total time of the simulation is assumed to be 60s, and the fourth-order Runge-Kutta method is used to discretized the continuous state equation with a sample frequency 10Hz. 200 Monte Carlos are run to evaluate the performance of the three filters. In the imperfect CKF and the DCKF, the uncertain parameter $c$ is set as the corresponding reference value, $\bar{c} = 2\times 10^4$. For the DCKF, the sensitivity-weighting matrix is set to

$$W = diag\{3\times 10^4, 6\times 10^3, 1\times 10^5\} \quad (55)$$

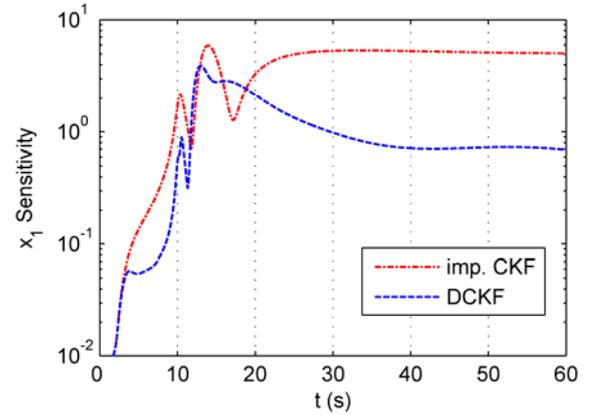

Fig.1. Sensitivities of the imperfect CKF and the DCKF for state $x_1$

The state sensitivities respected to the uncertain parameter $c$ with logarithmic scales are shown in Figs. 1 to 3. Compared to the imperfect CKF, the proposed DCKF almost has smaller sensitivities to the uncertain parameter. Figures 4 to 5 show that the root mean squared errors (RMSEs) of the above three filters. When the process model is disturbed by the uncertain parameter, the RMSE for the DCKF is smaller than that of the imperfect CKF. In a word, the proposed DCKF can reduce the sensitivities of the state estimate errors respect to the uncertain parameter compared with the imperfect CKF, and has a better performance than the imperfect CKF when the process model parameter is imperfect.

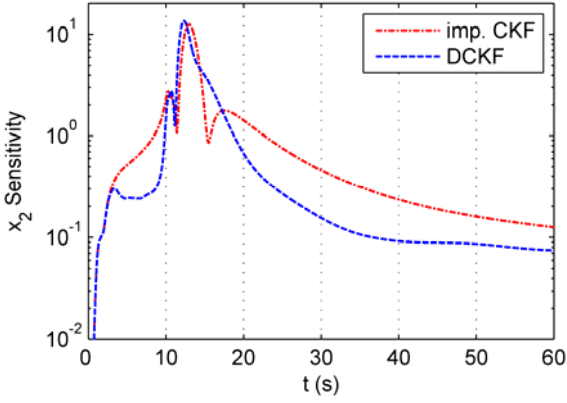

Fig.2. Sensitivities of the imperfect CKF and the DCKF for state $x_2$

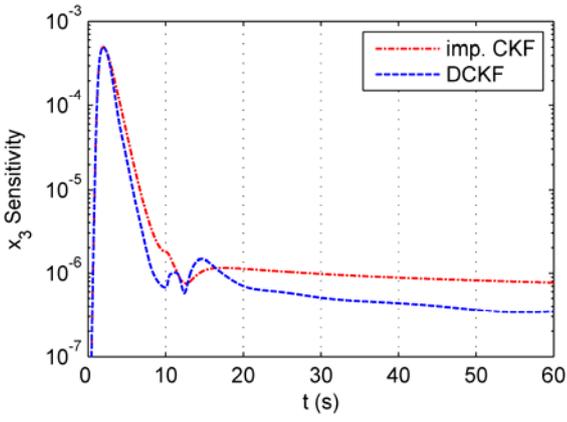

Fig.3. Sensitivities of the imperfect CKF and the DCKF for state $x_3$

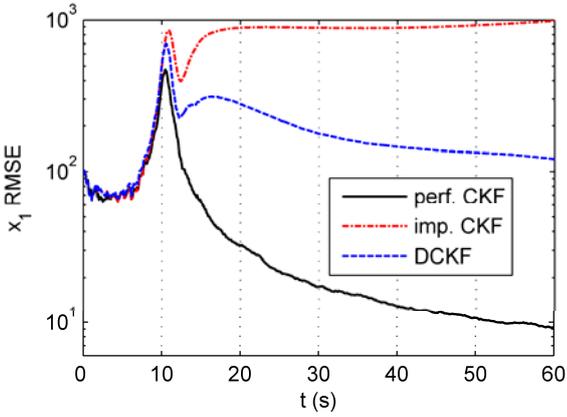

Fig.4. $x_1$ RMSE

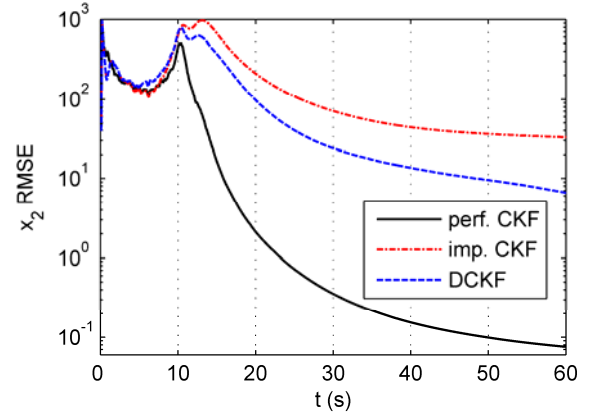

Fig.5. $x_2$ RMSE

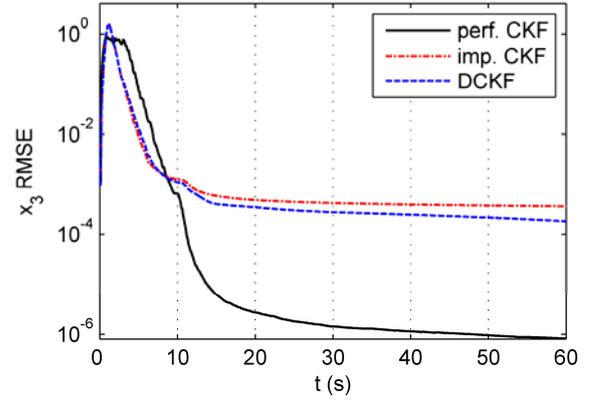

Fig.6. $x_3$ RMSE

### 4.2 Hovering helicopter model with two uncertain parameters

In this example, the hovering helicopter model is given by [21]

$$\dot{x} = \begin{bmatrix} c_1 & c_2 & -g & 0 \\ 1.26 & -1.765 & 0 & 0 \\ 0 & 1 & 0 & 0 \\ 1 & 0 & 0 & 0 \end{bmatrix} x - \begin{bmatrix} 0.086 \\ -7.408 \\ 0 \\ 0 \end{bmatrix} K_{lqr} x \quad (56)$$

$$z_k = x_k + v_k \quad (57)$$

where the state vector is $x = [x_1, x_2, x_3, x_4]^T$, in which $x_1$ is the horizontal velocity, $x_2$ is the pitch angle of the fuselage and its derivative $x_3$, and $x_4$ is perturbation from a ground point reference. $g$ is the acceleration from gravity and its value is $g = 0.322$, $K_{lqr}$ is a constant vector given by $K_{lqr} = [1.989, -0.256, -0.7589, 1]$. $v_k$ is the measurement noise. The two uncertain model parameters $c_1$ and $c_2$ are assumed to be uniform distributions, which are respectively $c_1 \square U(-0.15, -0.05)$ and ... The initial true state of the HHM is

$$x_0 = [0.7929, -0.0466, -0.1871, 0.5780]^T \quad (58)$$

and the initial state of the filter is set as $\hat{x}_0 = x_0$, with the initial covariance $\hat{P}_0 = I$.

In simulation, the total time interval is assumed to be 4s, and the continuous state equation has been discretized using the fourth-order Runge-Kutta method with a sample frequency 20Hz. For performance evaluation, 200 Monte Carlo runs are performed for three filters. In the imperfect CKF and the DCKF the two uncertain parameters are set as their reference values, which are $\bar{c}_1 = -0.1$ and $\bar{c}_2 = 0.1$. For the DCKF, the sensitivity-weighting matrices are set to

$$W_1 = W_2 = diag\{3\times10^{-3}, 2\times10^{-3}, 1\times10^{-2}, 2\times10^{-2}\} \quad (59)$$

The true state and its estimates of three filters, RMSE, sensitivities respect to the uncertain parameters, and mean cost functions of the three filters are performed in the following.

The true state of the HHM and its estimates of the imperfect CKF and the DCKF in one Monte Carlo test are shown in Fig. 7. The true values of the two parameters in this simulation case are $\bar{c}_1 = -0.0729$ and $\bar{c}_2 = 0.1142$, respectively.

Figure 8 represents the state sensitivities respect to the uncertain parameter, $c_1$, in the imperfect CKF and the DCKF with logarithmic scales, because the state sensitivities to $c_2$ are similar. From Fig. 8, it can be seen that the imperfect CKF almost has a larger sensitivities to the parameter uncertainties than the DCKF. The mean cost functions of three filters with logarithmic scales are shown in Fig. 9. Here, for the different cost functions of the three filters, the cost function of the perfect CKF is $J_c = Tr(P_k^+)$, and the cost function of the imperfect CKF and the DCKF is computed using (49). It can be seen that the mean cost function of the perfect CKF is the smallest; the mean cost function of the DCKF takes second place, and the mean cost function of the imperfect CKF is the largest one. So, the DCKF reduces the sensitivities of the state estimate errors respect to the parameter uncertainties, and minimizes the corresponding cost function compared with the imperfect CKF.

Figure 10 shows that the RMSEs of the three filter. The RMSE for the DCKF is larger than the perfect CKF, and smaller than the imperfect CKF, when the process model has the uncertain parameters. These results demonstrate that the DCKF can effectively reduce the estimation errors in the presence of the uncertain parameters.

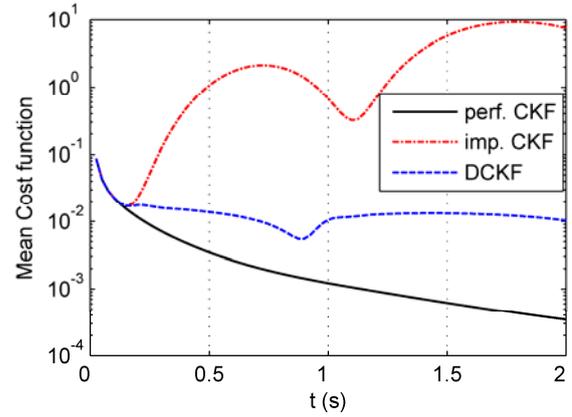

Fig.9. Mean cost function

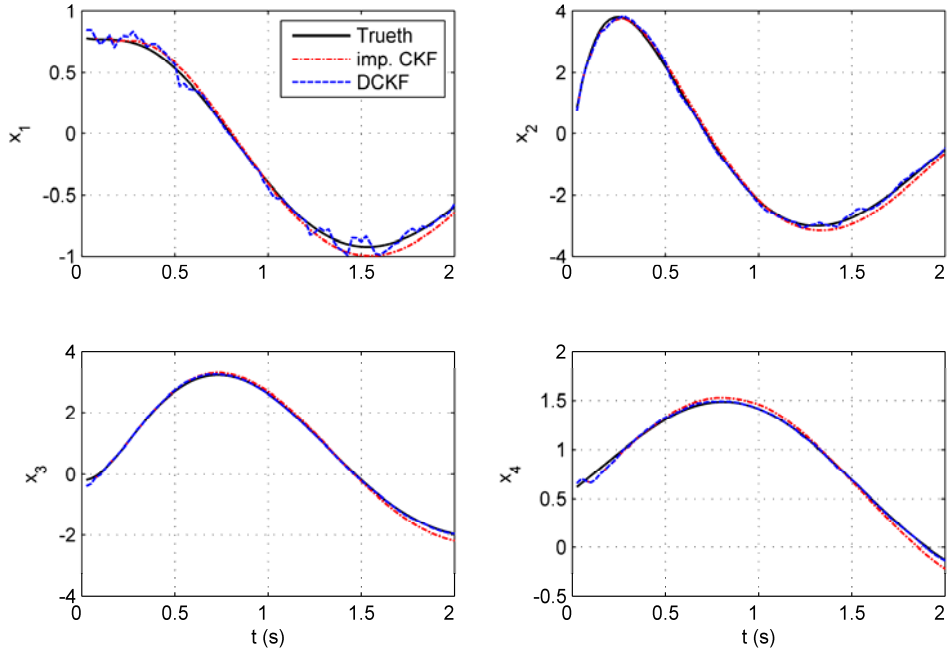

Fig.7. True values of the HHM and estimates of the imperfect CKF and the DCKF

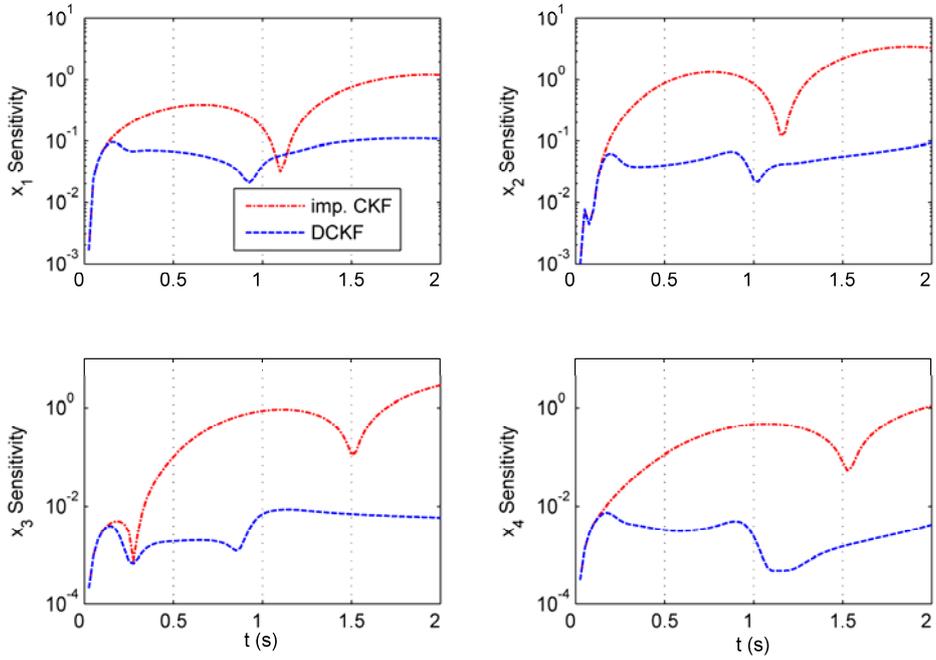

Fig.8. Sensitivities of the imperfect CKF and the DCKF

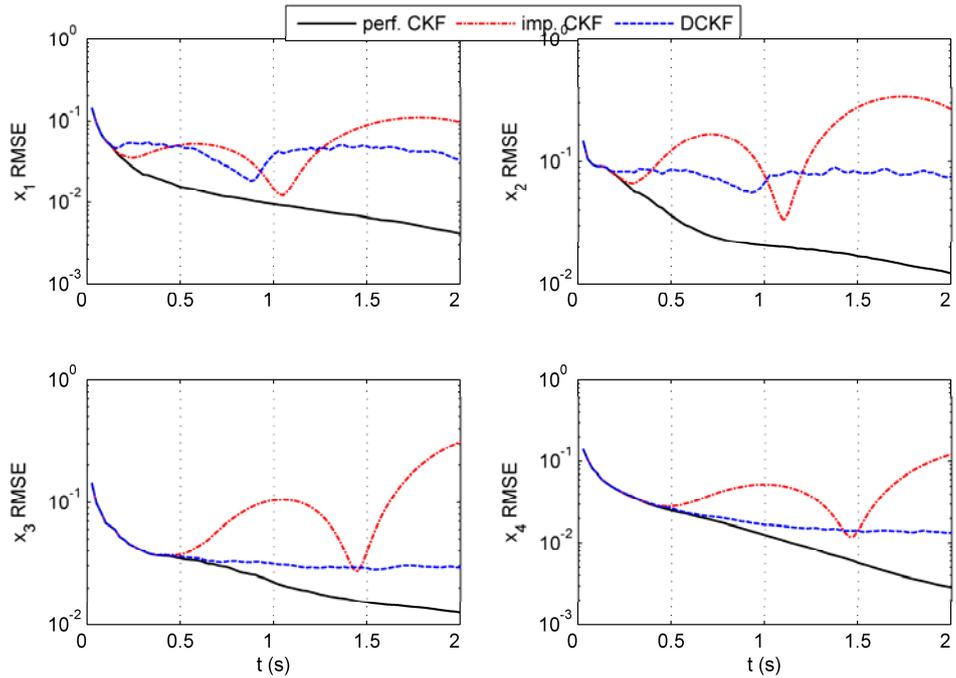

Fig.10. RMSE

A filter consistency test based on the normalized mean error (NME) test described in reference is introduced to highlight the effectiveness of the covariance estimates in accounting for errors in the state estimates. The results of the NME consistency test are shown in Fig. 11. It can be seen that all the state test statistics of the proposed DCKF are well below the required threshold, and this shows that the DCKF covariance estimate accurately predicts the state estimate errors. But, the results of the imperfect CKF indicate that it has a poor consistency in describing the states. That is to say, the DCKF consistently provides excellent state estimates when the process model parameter has uncertainties.

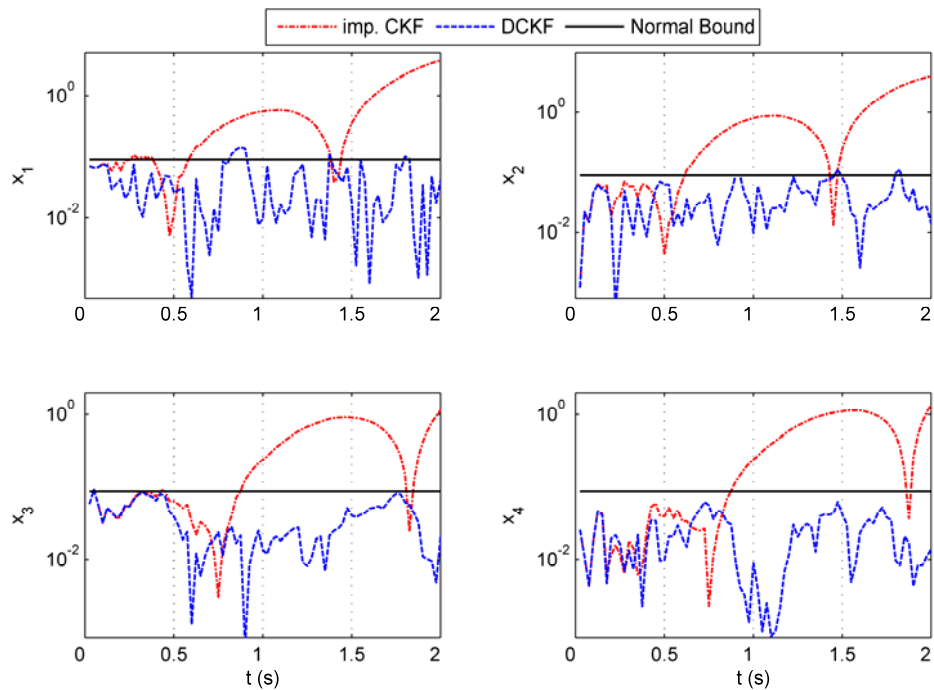

Fig.11. NME consistency test

## 5. CONCLUSIONS

In this paper, the state estimation of nonlinear systems with uncertain parameter is studied based on the desensitized optimal control methodology. The definition of the state estimate error sensitivity to the uncertain parameter is introduced into the recursive Bayesian filter. The robust desensitized cubature Kalman filtering (DCKF) for nonlinear systems with uncertain parameter under Bayesian framework is proposed. The desensitized cost function of the DCKF is designed by penalizing the posterior covariance trace by a sensitivity-weighting sum of the a posteriori sensitivities, and the gain of the DCKF is obtained by minimizing the desensitized cost function to amend the state estimation. In the DCKF, the sensitivity propagation of the state estimate errors is described, and a special equation, which is different from the equation of Shen and Karlgaard, is solved to obtain the sensitivity of the root square matrix. Two numerical simulations are computed to verify the effectiveness of the proposed DCKF.